# Phase stabilization of In$_2$Se$_3$ by disordered Ni intercalation and its enhanced thermoelectrical performance

Zengguang Shi[1,2#], Yukun Xiao[2,3#], Mian Li[2,3*], Jianfeng Cai[2], Yanmei Chen[2,3], Jun Jiang[2], Xiaoping Ouyang[4], Zhifang Chai[2,3], Qing Huang[2,3*]

[1] School of Materials Science and Chemical Engineering, Ningbo University, Ningbo, Zhejiang 315211, China

[2] Ningbo Institute of Materials Technology and Engineering Chinese Academy of Sciences, Ningbo 315201, China

[3] Qianwan Institute of CNITECH, Ningbo 315336, China.

[4] School of Materials Science and Engineering, Xiangtan University, Xiangtan 411105, China

E-mail addresses: limian@nimte.ac.cn (M. Li), huangqing@nimte.ac.cn (Q. Huang)

## Abstract

Van der Waals (*vdW*) layered materials have gained significant attention owing to their distinctive structure and unique properties. The weak interlayer bonding in vdW layered materials enables guest atom intercalation, allowing precise tuning of their physical and chemical properties. In this work, a ternary compound, Ni$_x$In$_2$Se$_3$ (x = 0–0.3), with Ni randomly occupying the interlayers of In$_2$Se$_3$, was synthesized via an intercalation route driven by electron injection. The intercalated Ni atoms act as "anchor points" within the interlayer of In$_2$Se$_3$, which effectively suppresses the phase transition of In$_2$Se$_3$ at elevated temperatures. Furthermore, the disordered Ni intercalation significantly enhanced the electrical conductivity of In$_2$Se$_3$ through electron injection, while reducing the thermal conductivity due to the interlayer phonon scattering, leading to an improved thermoelectric performance. For instance, the thermoelectric figure of merit (*ZT*) of Ni$_{0.3}$In$_2$Se$_3$ increased by 86% (in-plane) and 222% (out-of-plane) compared to In$_2$Se$_3$ at 500 °C. These findings not only provide an effective strategy to enhance the performance of layered thermoelectric materials, but also demonstrate the potential of intercalation chemistry for expanding the application scope of van der Waals (*vdW*) layered materials.

## Introduction

Van der Waals (*vdW*) layered materials, such as graphene, TMDs, and MXenes[1,2], exhibit tremendous potential in energy storage[3], catalysis[4], sensing[5], and other fields due to their distinctive layered structures and unique properties. The *vdW* layered materials are usually combined with strong intra-layer bonding and weak inter-layer interactions, which provides the possibility for guest substances intercalating into the *vdW* gap. The electronic structure, interlayer spacing, and physicochemical properties of *vdW* layered materials can be effectively regulate the by inserting atoms, molecules, or ions into the *vdW* gap, and thereby expanding their application scope[6,7]. To date, researchers have developed diverse intercalation methods including electrochemical intercalation, chemical intercalation, and ion exchange. These techniques have enabled successful insertion of various guest species into different layered materials, allowing for precise control of material properties[8]. For example, Gong et al. realized Cu and Co intercalation into SnS$_2$ though liquid-phase reaction, resulting in the transformation of the pristine n-type semiconductor SnS$_2$ into p-type Cu-SnS$_2$ and metallic Co-SnS$_2$[9]. Chen et al. achieved Co intercalation into MoS$_2$ through an electrochemical route, which results in chemo-selective oxidation of sulfides to sulfones and enhanced catalytic performance[10]. Li et al. used liquid-phase intercalation to insert K$^+$ into MXene layers, which significantly improved the electrical performance of MXene as supercapacitors[11].

As a *vdW* layered material, In$_2$Se$_3$ has attracted considerable attention in recent years due to its potential applications in nanoelectronics, optoelectronics, thermoelectrics, non-volatile memories, and gas sensors[12]. In$_2$Se$_3$ exhibits multiple crystal structures, including α, β, γ, and κ phases[13], each demonstrating distinct physical and chemical properties. For instance, α-$In_2Se_3$ is a ferroelectric material[14], while β-$In_2Se_3$ exhibits semiconductor characteristics[15]. The structural and phase diversity of $In_2Se_3$ offers extensive opportunities for property tuning, but it also poses challenges for practical applications. $In_2Se_3$ undergoes a serials of reversible phase transition above 100 °C, which cause a dramatic change in properties such as electronic transport properties, electrical conductivity and Optical properties[16]. Such temperature-dependent phase transitions constrain its utility in thermally variable environments. Thus, extensive research has been devoted to controlling the phase transition of $In_2Se_3$. Meanwhile, enhancing the performance of $In_2Se_3$ for practical applications through structural and compositional optimization remains a key research focus. For instance, Okmin Park[17] et al. found that Si doping can reduce the thermal conductivity of $In_2Se_3$, thus results in a improved thermoelectric performance. However, so far, most studies on modifying the properties of $In_2Se_3$ have focused on heteroatom doping, while property modulation *via* intercalation remains rare.

In this work, intercalation strategy is proposed to regulate the structural and tailor the properties of $In_2Se_3$. A ternary compound, $Ni_xIn_2Se_3$ (x = 0–0.3), with Ni randomly occupying the interlayers of $In_2Se_3$, was synthesized through an "chemical scissor" intercalation route. The intercalation of Ni significantly increased the carrier concentration, leading to enhanced electrical conductivity. Meanwhile, the disordered intercalation structure strengthened phonon scattering, resulting in reduced thermal conductivity. These synergistic effects collectively improved the thermoelectric performance of the layered compounds. The thermoelectric figure of merit (*ZT*) of $Ni_{0.3}In_2Se_3$ increased by 86% (in-plane) and 222% (out-of-plane) compared to $In_2Se_3$ at 500 °C. In addition, Ni intercalation effectively suppresses the phase transition of $In_2Se_3$ at elevated temperatures, leading to improved thermal stability and linear temperature dependence of physical properties of the intercalation compounds. These findings not only establish an effective strategy for enhancing the performance of layered thermoelectric materials, but also highlight the potential of intercalation chemistry for broadening the application scope of *vdW* layered materials.

# Experimental

## Preparation of In$_2$Se$_3$ and Ni$_x$In$_2$Se$_3$ powders

In$_2$Se$_3$ was synthesized through a solid-state reaction. Indium (In) and selenium (Se) powders in a 2:3 molar ratio were thoroughly mixed in a mortar. The mixture was then loaded into a quartz tube, which was then evacuated into vacuum and flame-sealed using a H$_2$/O$_2$ torch. The sealed tube was placed in a muffle furnace, heated to 900 °C with the rate of 4 °C/min, and held at this temperature for 10 hours before natural cooling to room temperature. As the result, In$_2$Se$_3$ powders were obtained in the quartz tube.

Ni$_x$In$_2$Se$_3$ was synthesized through an intercalation reaction in LiCl-KCl eutectic molten salt. Typically, a mixture of In$_2$Se$_3$, Ni powder, KCl, and LiCl in a 1:x:10:15 molar ratio was thoroughly mixed in a mortar under argon protection. The mixture was then transferred to an alumina crucible and loaded into a tube furnace. The furnace was heated to 450 °C at a rate of 4 °C/min and maintained at this temperature for 5 hours under argon flow. After natural cooling to room temperature, the product was washed repeatedly with deionized water to remove residual LiCl and KCl, followed by drying at 60 °C for 2 hours to obtain final Ni$_x$In$_2$Se$_3$ powders.

## Preparation of In$_2$Se$_3$ and Ni$_{0.3}$In$_2$Se$_3$ bulks

To investigate the properties of In$_2$Se$_3$ and Ni$_{0.3}$In$_2$Se$_3$, bulk samples were fabricated using spark plasma sintering (SPS). Taking Ni$_{0.3}$In$_2$Se$_3$ as an example, the synthesized Ni$_{0.3}$In$_2$Se$_3$ powder was loaded into a 20 mm diameter graphite die. The sintering process was performed under an argon atmosphere in an SPS system. The graphite die was heated to 500 °C at a rate of 50 °C/min and held at this temperature under 40 MPa pressure for 10 minutes, yielding dense Ni$_{0.3}$In$_2$Se$_3$ bulk sample. The In$_2$Se$_3$ bulk sample was prepared following the same sintering procedure.

The In powder and Ni powder with a particle size of 300 mesh were purchased from Pantian Nanomaterials Co. Ltd. (Shanghai, China). Se powder with the particle size of 300 mesh, KCl and LiCl, were acquired from Aladdin Industrial Co., Ltd. (Shanghai, China).

**Characterization**

The crystal structure and composition of the samples were analyzed at 40 KV and 40 mA using a D8 Advance X-ray diffractometer manufactured by Bruker AG in Germany. The topography of the samples was observed using a G300 thermal field scanning electron microscope from ZEISS in Germany and a Hitachi 8230 model cold field scanning electron microscope from Japan, both equipped with Bruker X-ray energy dispersive analyzers. An AXIS SUPRA+ X-ray Photoelectron Spectrometer (AXIS SUPRA+) from the manufacturer Shimadzu, Japan, was used to analyze the composition and chemical values of the samples. The atomic structure of the materials was characterized using a Spectra 300 model double spherical aberration corrected projection electron microscope manufactured by Thermo Fisher, USA. The thermal conductivity of the samples was measured using a Laser Thermal Conductivity Meter (LFA 467) from NETZSCH, Germany, the electrochemical properties of the samples were measured using a Hall Effect Test System (8404-CRX-6.5K) from LakeShore, U.S.A., and the Seebeck coefficients and conductivities of the samples were measured using a ZEM-3 system from Ulbac-Riko, Japan. The bulk $In_2Se_3$ and $Ni_{0.3}In_2Se_3$ samples were precision-cut into two different dimensions: 10×10×2 mm specimens for thermal and electrical characterization, and 12×2×2 mm specimens for thermoelectric measurements.

# Results and Discussion

# 1. Materials Characterization

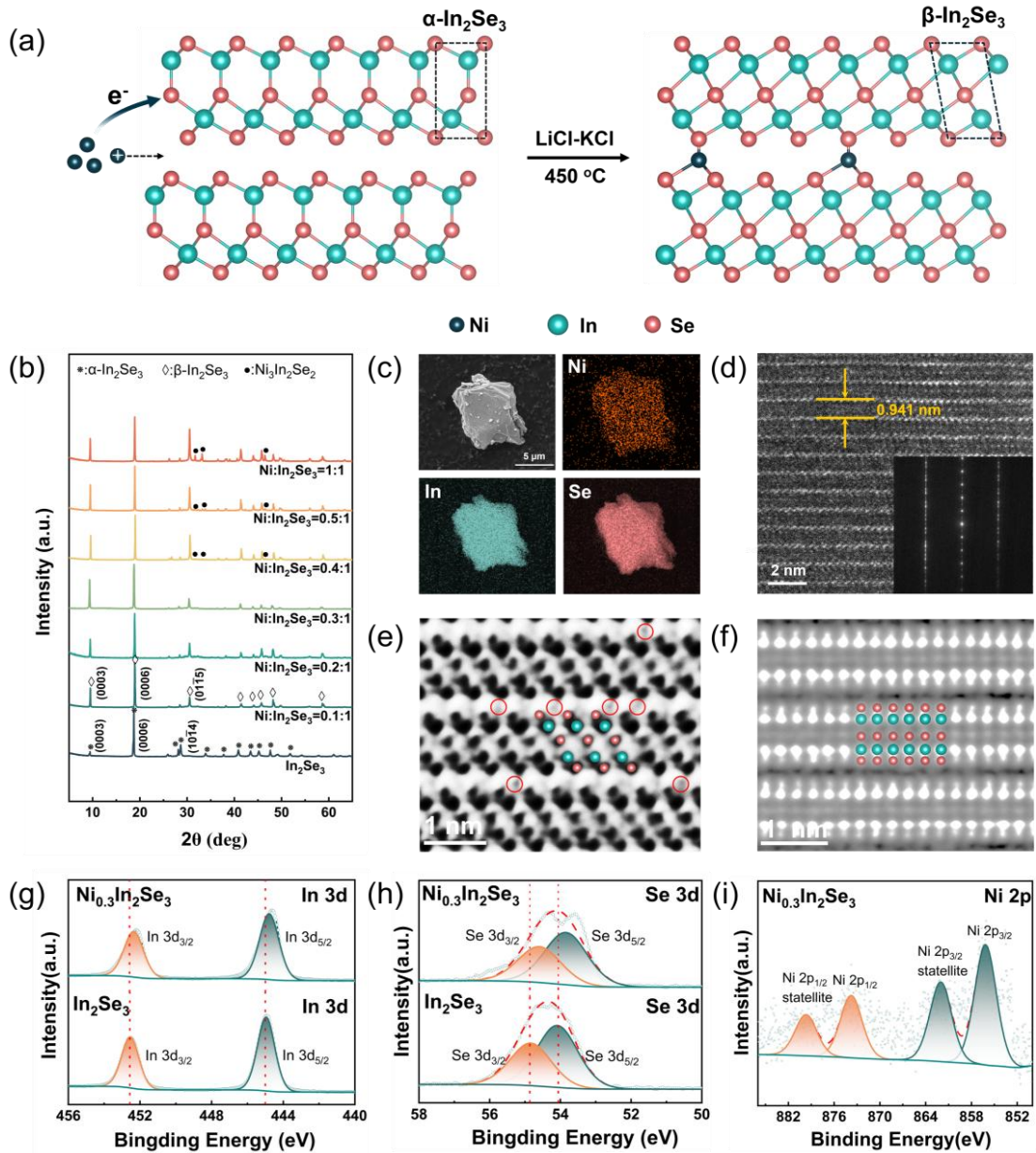

Fig .1 (a) Schematic illustration of the intercalation process. (b) XRD patterns of In$_2$Se$_3$ with different Ni intercalation concentrations. (c) SEM image and EDS maps showing a particle of Ni$_{0.3}$In$_2$Se$_3$. (d) HR-TEM image and SAED pattern of Ni$_{0.3}$In$_2$Se$_3$ along the [000$l$] axis. (e) BF-STEM images of Ni$_{0.3}$In$_2$Se$_3$ along [11$\bar{2}$0] axis. (f) HAADF-STEM images of Ni$_{0.3}$In$_2$Se$_3$ along [1$\bar{1}$00] axis. (g) In 3d XPS spectra of Ni$_{0.3}$In$_2$Se$_3$ and In$_2$Se$_3$. (h) Se 3d XPS spectra of Ni$_{0.3}$In$_2$Se$_3$ and In$_2$Se$_3$. (i) Ni 2p XPS spectra of Ni$_{0.3}$In$_2$Se$_3$.

Fig. 1a illustrates the intercalation process. When metal Ni is added in the molten salts, the polarization effect of the molten salts causes the separation of valence electrons from the Ni atoms[18]. These electrons are subsequently captured by In$_2$Se$_3$, leading to negatively charged In$_2$Se$_3$ layers. Concurrently, Ni$^+$ cations spontaneously intercalate into these negatively charged *vdW* gaps of $In_2Se_3$, forming the intercalation compound. This electron injection and intercalation process following the "chemical scissor" intercalation protocol in our previous reports, which used solvated metals as reductive chemical scissors to tailor the surface termination of MXenes and intercalate transition metal dichalcogenides with metal atom[19,20].

Fig. 1b present the XRD patterns of $In_2Se_3$ before and after intercalation reaction. The pristine $In_2Se_3$ exists in the α-phase with the space group of *R3m*, which can be confirmed by comparing with standard PDF card (97-001-7008). After intercalation reaction, the intensity of (0003) and ($01\bar{1}5$) peaks increased significantly. This observation can be attributed to the phase transformation from α-$In_2Se_3$ to β-$In_2Se_3$ (space group *R3m*). The α→β phase transition involves reversible intralayer atomic plane slippage, with the corresponding atomic structures of both α-$In_2Se_3$ and β-$In_2Se_3$ phases illustrated in Fig. 1a. Additionally, in samples with higher Ni concentrations (e.g, Ni:$In_2Se_3$ = 0.4:1, 0.5:1 and 1:1), a secondary phase of $Ni_3In_2Se_2$ was detected in the reaction products. This observation indicates that the Ni content exceeds the maximum intercalation capacity. To determine this maximum intercalation concentration, we employed EDS analysis to characterize the reaction products of Ni-$In_2Se_3$ samples with varying Ni ratios. EDS analysis of high Ni-ratio samples (Ni:$In_2Se_3$ > 0.3:1) revealed an actual Ni:$In_2Se_3$ ratio of approximately 0.3:1 within intact $In_2Se_3$ particles. The SEM and EDS analyses shown in Figs. S1-S6 and Table S1 confirm this observation. This result agrees with XRD observations: when the Ni ratio exceeds 0.3 in the starting materials, a second phase ($Ni_3In_2Se_2$) appears in the reaction products. As shown in Fig. S7 and Table S2, the by-product $Ni_3In_2Se_2$ are irregular particles with high Ni content. Thus, the sample with maximum Ni intercalation concentration is determined as $Ni_{0.3}In_2Se_3$. Fig. S8 displays the Rietveld refinement pattern of the $Ni_{0.3}In_2Se_3$ sample, confirming the preservation of a hexagonal crystal structure *R3m* with refined lattice parameters of $a$ = 3.973 Å and $c$ = 28.31 Å. The corresponding atomic positions are tabulated in Table S3. Note that the $c$ lattice parameter of $Ni_{0.3}In_2Se_3$ is smaller than that of pristine $In_2Se_3$ (28.75 Å), suggesting stronger interlayer bonding. Additionally, we have tried to synthesize $Ni_{0.3}In_2Se_3$ directly through solid-solid reaction by using Ni, In and Se as the raw materials. As shown in the results of Fig. S9, single phase $Ni_{0.3}In_2Se_3$ cannot be synthesized by the solid-phase reaction because the formation of competing phases such as $NiSe$, $In_6Se_7$.

Fig. 1c presents the scanning electron microscopy (SEM) image and corresponding energy-dispersive X-ray spectroscopy (EDS) elemental maps of a $Ni_{0.3}In_2Se_3$ particle. The $Ni_{0.3}In_2Se_3$ maintains morphological similarity to pristine $In_2Se_3$, demonstrating that the intercalation process preserves the structural integrity. EDS mapping reveals homogeneous elemental distribution, and quantitative analysis in Fig. S3 yields Ni:In:Se = 5.5:36.4:58.4 (at%), which is consistent with the expected stoichiometric ratio of $Ni_{0.3}In_2Se_3$.

Fig. 1d presents the high-resolution transmission electron microscopy (HRTEM) image and corresponding selected area electron diffraction (SAED) of $Ni_{0.3}In_2Se_3$ along the [000$l$] axis. The d-spacing value of 0.941 nm corresponds to the (0003) plane, which is consistent with the XRD refinement results. The SAED pattern exhibits non-uniform intensity distribution of diffraction spots. Specifically, diffraction spots along the same streak show varying brightness levels, with some diffraction streaks appearing sharp while others appear relatively diffuse. This phenomenon can be attributed to the disordered intercalation of Ni atoms within the interlayers of the $In_2Se_3$ material, resulting in differential electron scattering from various regions of the sample[21]. The disordered arrangement of intercalated Ni atoms disrupts the periodicity of the pristine crystal structure, causing regional variations in the diffraction patterns.

Fig. 1e presents a bright-field scanning transmission electron microscopy (BF-STEM) image of $Ni_{0.3}In_2Se_3$ captured along the [11$\bar{2}$0] zone axis. The image confirms that $Ni_{0.3}In_2Se_3$ stabilizes in the β-phase structure following intercalation. Notably, the black holes highlighted by red circles represent Ni atoms. The intercalated Ni atoms are observed to randomly occupy tetrahedral sites within the $In_2Se_3$ interlayer. Fig. 1f and Fig. S10 provides high-angle annular dark-field scanning transmission electron microscopy (HAADF-STEM) images and corresponding EDS line scanning of

$Ni_{0.3}In_2Se_3$ along the [1$\bar{1}$00] direction, which provide further evidence for the disordered arrangement of Ni atoms within the $In_2Se_3$ interlayers.

X-ray photoelectron spectroscopy (XPS) analysis was employed to investigate the electronic transfer behavior of the intercalation reaction. Figs. 1g-1i present comparative XPS characterization of samples before and after intercalation. Fig 1g displays the In *3d* spectra, which exhibit a bimodal structure due to spin-orbit coupling. These peaks are identified as In $3d_{3/2}$ (452.56 eV) and In $3d_{5/2}$ (445 eV). After Ni intercalation, the In $3d_{3/2}$ and In $3d_{5/2}$ peaks shift toward lower binding energy by 0.19 eV, indicating a decreased oxidation state valence. This phenomenon occurs because the intercalated Ni contributes electrons to $In_2Se_3$, as described in the reaction process. Fig. 1h depicts the Se 3d spectra. The binding energy peaks of Se are also shifted to lower binding energy peaks after Ni intercalation, which can be attributed to the formation of Ni-Se bonding. In addition, Ni intercalation stabilizes $In_2Se_3$ in its β-phase, which have different In-Se bond lengths and angles to that of pristine α-$In_2Se_3$. This fact in turn affects the electronic states and binding energies of Se. Fig 1i presents the Ni 2p XPS spectra of $Ni_{0.3}In_2Se_3$. The peaks at 856 eV and 874 eV correspond to Ni-Se ($2p_{3/2}$) and Ni-Se ($2p_{1/2}$) bonds, which confirms that Ni predominantly exists in the $Ni^{2+}$ oxidation state in $Ni_{0.3}In_2Se_3$[22].

## 2. Phase stability

Thermal analysis and in-situ high-temperature XRD were employed to investigate the thermal stability of both $In_2Se_3$ and $Ni_{0.3}In_2Se_3$. Fig. 2a presents the differential scanning calorimetry (DSC) curves for $In_2Se_3$ and $Ni_{0.3}In_2Se_3$. $In_2Se_3$ exhibits two exothermic peaks at approximately 100 °C and 200 °C, relating to its α-to-β and β-to-γ phase transition. In contrast, $Ni_{0.3}In_2Se_3$ maintains a relatively flat DSC curve throughout the entire temperature range with no discernible thermal effect peaks. Fig 2b displays thermogravimetric analysis (TGA) curves for both pristine $In_2Se_3$ and $Ni_{0.3}In_2Se_3$. The thermogravimetric data demonstrates that neither sample experiences measurable weight loss across the investigated temperature range, confirming that the materials do not undergo volatilization or decomposition processes that would result in mass reduction.

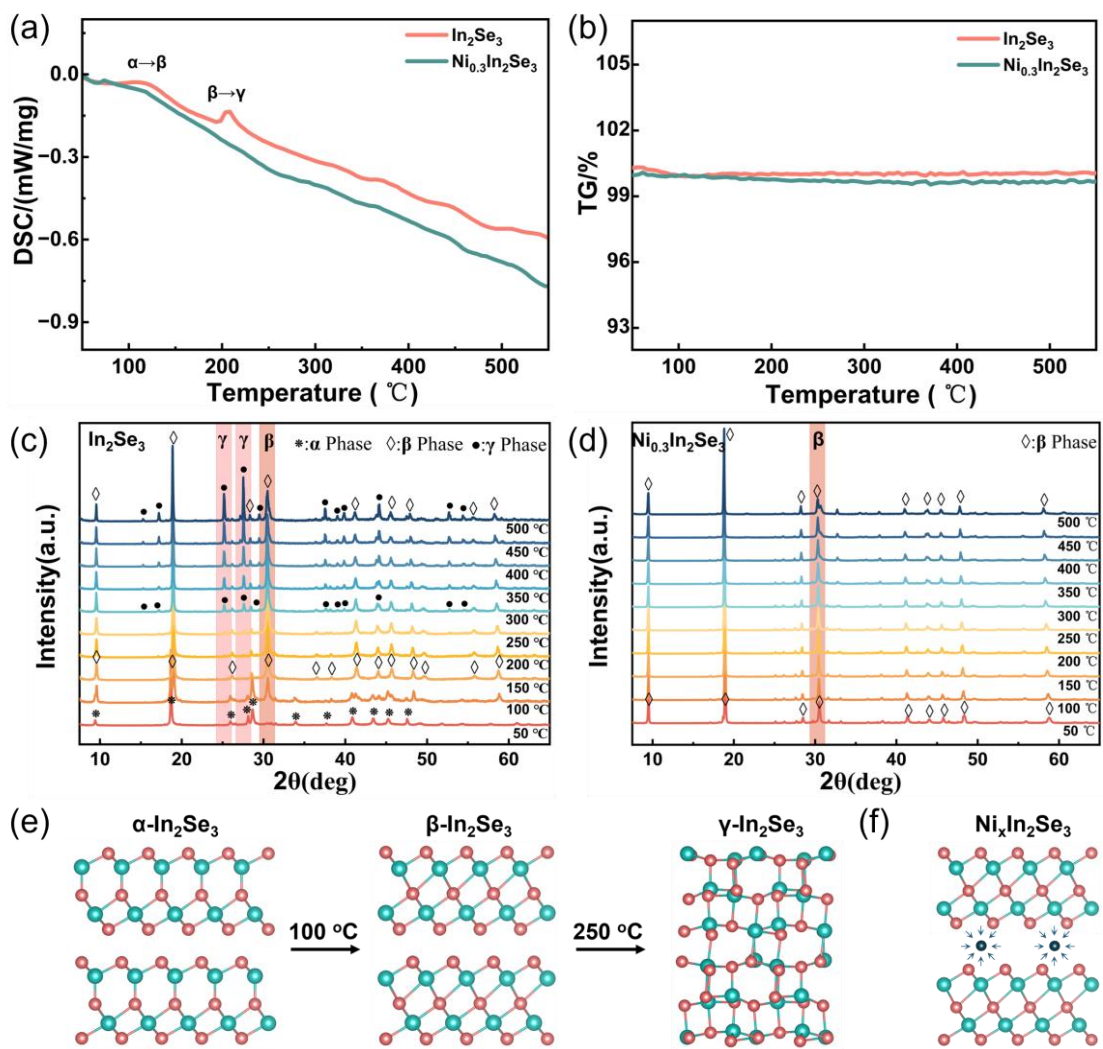

Fig 2. (a-b) DSC and TG curves of In$_2$Se$_3$ and Ni$_{0.3}$In$_2$Se$_3$ at the temperature range from 25 °C to 500 °C. (c-d) In-situ high-temperature XRD stacked patterns of pristine In$_2$Se$_3$ and Ni$_{0.3}$In$_2$Se$_3$. (e-f) the schematic illustration of the phase transition process of In$_2$Se$_3$ and Ni$_{0.3}$In$_2$Se$_3$.

Fig. 2c-2d display the in-situ high-temperature XRD stacked patterns of In$_2$Se$_3$ and Ni$_{0.3}$In$_2$Se$_3$. It is revealed that pristine In$_2$Se$_3$ undergo substantial phase transitions with increasing temperatures. At room temperature, pristine In$_2$Se$_3$ exists in the α-phase configuration. As temperature increases to 150 °C, a complete transformation from α-phase to β-phase occurs. Upon reaching 250 °C, the material begins transitioning from β-phase to γ-phase, as illustrated in Fig. 2e. These phase transitions agree well with the thermal effect peaks observed in Fig. 2a and account for the significant changes in the electronic transport properties of In$_2$Se$_3$ near 200 °C. In contrast, Ni$_{0.3}$In$_2$Se$_3$

demonstrates remarkable thermal stability throughout the entire measured temperature range. $Ni_{0.3}In_2Se_3$ stabilizes in the β-phase configuration, with neither the intensity nor the position of XRD diffraction peaks changing with temperature variations. The stabilization of the β-phase in $Ni_{0.3}In_2Se_3$ can be attributed to the anchoring effect of the intercalated Ni atoms. As demonstrated by Zhang[23] et al., the phase transition of $In_2Se_3$ follows an unusual intralayer-splitting/interlayer-reconstruction pathway, facilitated by the weak *vdW* forces between the layers. However, for $Ni_{0.3}In_2Se_3$, as revealed by XPS and XRD analysis, Ni forms strong bonds with Se, leading to a reduced interlayer spacing compared to $In_2Se_3$. These facts make the intercalated Ni atoms act as "anchor points", preventing the intralayer-splitting/interlayer-reconstruction required for phase transition (Fig. 2f). This fact is consistent with the report by Chu et al., which showed that the intercalation of $Fe^{3+}$ in layered $Na_{0.67}Mn_{0.5}Co_{0.3}Fe_{0.2}O_2$ can enhance its structural stability[24].

## 3. Physical properties

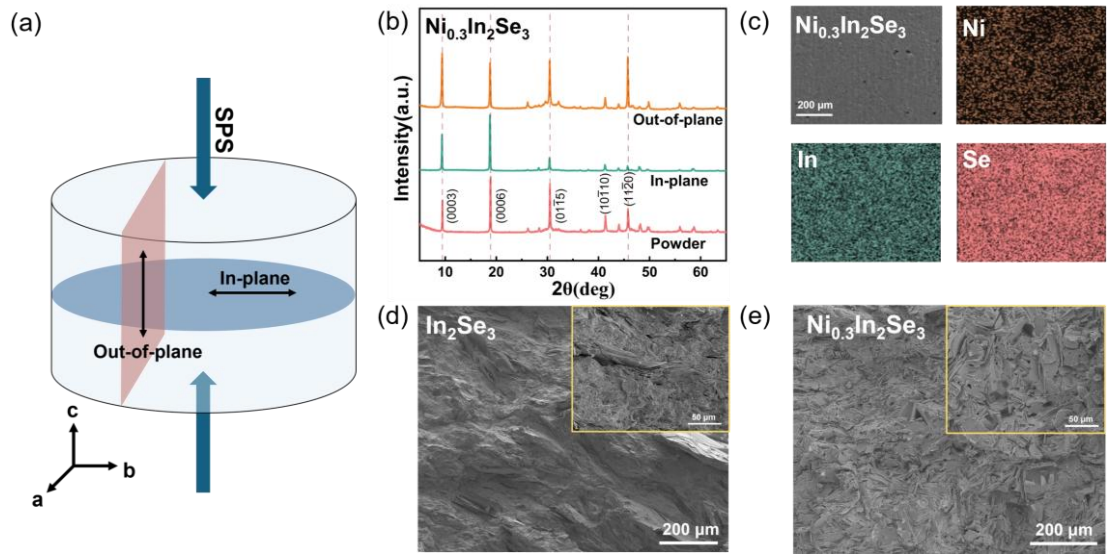

Fig 3. Characterization of bulk $In_2Se_3$ and $Ni_{0.3}In_2Se_3$ sample: (a) Schem of samples cutting for properties measurement. (b) XRD patterns of $Ni_{0.3}In_2Se_3$ powder and bulk simples cut along different directions. (c) SEM and EDS maps of the surface of $Ni_{0.3}In_2Se_3$ bulk. (d-e) SEM images of the fracture surface of $In_2Se_3$ bulk and $Ni_{0.3}In_2Se_3$ bulk.

To investigate the influence of Ni intercalation on material properties, bulk samples of $Ni_{0.3}In_2Se_3$ and $In_2Se_3$ were fabricated using spark plasma sintering (SPS). The as-obtained bulk samples are cylindrical in shape with a diameter of 20 mm and a height of approximately 13 mm (Figs. S11 and S12). The actual density of the $In_2Se_3$ bulk and $Ni_{0.3}In_2Se_3$ bulk is determined as 5.68g/cm$^3$ and 6.01g/cm$^3$, with the corresponding relative density of 98.6% and 96.1% respectively. The calculation details of the density are provided in SI (Table S4).

Considering the preferred orientation of the bulk samples, the bulk samples was cut into specimens along different directions to investigate their properties parallel to (out-of-plane) and perpendicular to (in-plane) the pressure, as illustrated in Fig. 3a. Fig 3b displays the XRD patterns of $Ni_{0.3}In_2Se_3$ bulk samples with different orientations. The results show that neither impurity phase formation nor phase transformation was detected in the samples during SPS, demonstrating good phase stability of the $Ni_{0.3}In_2Se_3$. Notably, the relatively intensity of (0003), (0006), (01$\bar{1}$5) and (11$\bar{2}$0) peaks of the samples cut in different orientations are significantly different, indicating the pressure-induced preferential orientation formation during sintering. The preferential orientation is also observed for $In_2Se_3$ bulk (Fig. S13). Fig. 3c presents the SEM image and corresponding EDS elemental maps of the $Ni_{0.3}In_2Se_3$ bulk surface, demonstrating a dense surface and uniform elemental distribution. Figs 3d and 3e shows the fracture surface of bulk $In_2Se_3$ and $Ni_{0.3}In_2Se_3$. The $In_2Se_3$ sample exhibits layered structure with clear preferred orientation caused by uniaxial pressure during the sintering process. $Ni_{0.3}In_2Se_3$ also shows a layered structure, while its preferred orientation is less obvious compared to $In_2Se_3$. This difference can be attributed to the intercalated Ni atoms functioning as pinning sites, which impede interlayer slippage and restrict the formation of highly oriented layers.

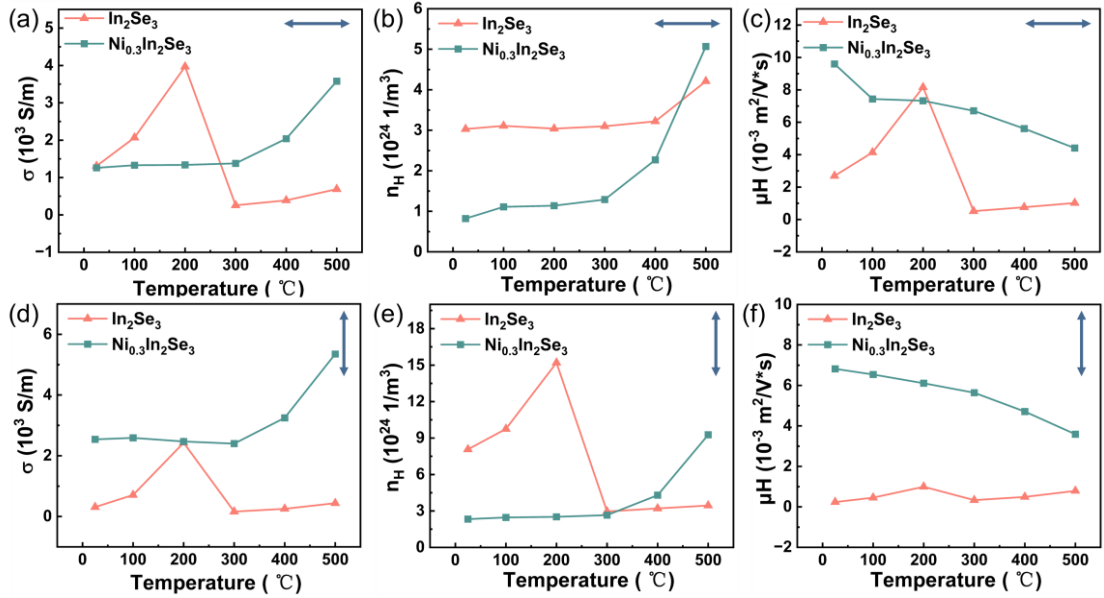

Fig. 4 (a-c) The in-plane direction electrical conductivity, carrier concentration, and carrier mobility of $In_2Se_3$ and $Ni_{0.3}In_2Se_3$. (d-f) The out-of-plane direction electrical conductivity, carrier concentration, and carrier mobility of $In_2Se_3$ and $Ni_{0.3}In_2Se_3$.

Fig. 4 presents a comparative analysis of electrical conductivity, carrier concentration, and carrier mobility for samples before and after Ni intercalation. Fig. 4a shows the temperature-dependent electrical conductivity of $In_2Se_3$ and $Ni_{0.3}In_2Se_3$ along the in-plane direction. The results show that compared to pristine $In_2Se_3$, the conductivity of $Ni_{0.3}In_2Se_3$ with significantly higher conductivity. The in-plane directional conductivity of $In_2Se_3$ is $1.31×10^3$ S/m at room temperature, and reaches the maximum value of $3.97×10^3$ S/m at 200 °C, and then decreases to $6.90×10^2$ S/m at 500 °C. In contrast, the in-plane directional conductivity of $Ni_{0.3}In_2Se_3$ as an in-plane directional conductivity of $1.26×10^3$ S/m at room temperature, which increases to $3.58×10^3$ S/m at 500 °C with increasing temperature. The increase in electrical conductivity of $Ni_{0.3}In_2Se_3$ can be attributed to the improvements in both carrier concentration and Hall mobility, as is illustrated in Figs 4b and 4c. At lower temperatures, the carrier concentration of $Ni_{0.3}In_2Se_3$ is $8.23×10^{23}$ $1/m^3$, which is lower than the $3.03×10^{24}$ $1/m^3$ of pristine in $In_2Se_3$. This reduction stems from deep energy level defects introduced during Ni intercalation[25], which effectively trap free electrons at low temperatures, thereby decreasing the overall carrier concentration. However, at elevated temperatures (above 300 °C), carriers trapped at these deep energy levels acquire sufficient thermal energy to achieve thermally activated delocalization, resulting in a rapid increase in carrier concentration. At 500 °C, the $Ni_{0.3}In_2Se_3$ carrier concentration is $5.07×10^{24}$ $1/m^3$, which is higher than the $4.21×10^{24}$ $1/m^3$ of $In_2Se_3$. Additionally, the intercalation did not change the carrier type - both pristine $In_2Se_3$ and $Ni_{0.3}In_2Se_3$ are N-type conductivity. The enhanced Hall mobility observed in $Ni_{0.3}In_2Se_3$ compared to pristine $In_2Se_3$ can be attributed to two primary factors. First, the intercalated Ni forms chemical bonds with Se, enhancing lattice stability and suppressing the formation of intrinsic defects at high temperatures. Second, Ni intercalation stabilizes the $In_2Se_3$ crystal structure in the β-phase, which offers more efficient carrier transport pathways than the α-phase. Together, these effects contribute to a significant improvement in carrier mobility across the entire measured temperature range.

The out-of-plane electrical conductivity, carrier concentration, and carrier mobility of $In_2Se_3$ and $Ni_{0.3}In_2Se_3$ are shown in Fig. 4d–4f. Their trends are generally similar to those in the in-plane direction. However, $Ni_{0.3}In_2Se_3$ exhibits reduced anisotropy compared to $In_2Se_3$, which can be attributed to its weaker preferred orientation, as discussed earlier.

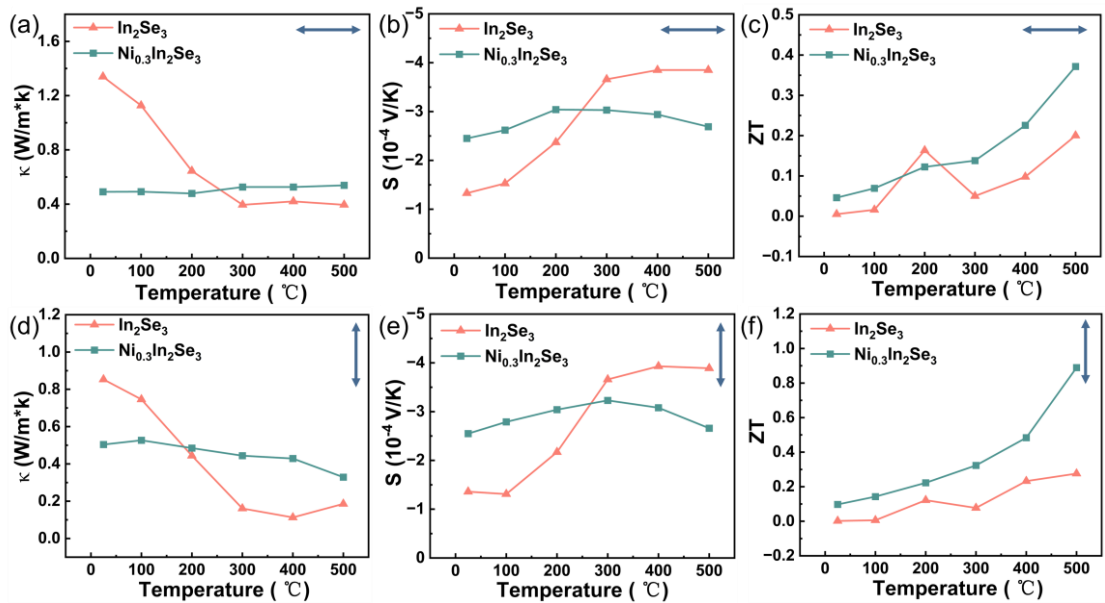

Fig. 5. (a-c) The in-plane direction thermal conductivity, Seebeck coefficient and thermoelectric merit *ZT* of $In_2Se_3$ and $Ni_{0.3}In_2Se_3$. (d-f) The out-of-plane direction thermal conductivity, Seebeck coefficient and thermoelectric merit *ZT* of $In_2Se_3$ and $Ni_{0.3}In_2Se_3$.

Figs. 5a and 5d present the temperature-dependent thermal conductivity of $In_2Se_3$ and $Ni_{0.3}In_2Se_3$. The in-plane thermal conductivity of $In_2Se_3$ is 1.339 W/m·K at room temperature and decreases significantly with increasing temperature, dropping to 0.395 W/m·K at 500 °C. In contrast, $Ni_{0.3}In_2Se_3$ exhibits much more stable thermal conductivity across the temperature range, maintaining 0.491 W/m·K at room temperature and 0.539 W/m·K at 500 °C. The out-of-plane thermal conductivities of both materials show similar temperature-dependent trends to their respective in-plane values. The reduced thermal conductivity of $Ni_{0.3}In_2Se_3$ compared to pristine $In_2Se_3$ can be attributed to modified phonon scattering mechanisms. In pure $In_2Se_3$, thermal transport is primarily dominated by intrinsic phonon-phonon scattering, which leads to strong temperature dependence of thermal conductivity. The introduction of Ni intercalation in $Ni_{0.3}In_2Se_3$ creates additional scattering pathways through two main effects: First, the randomly distributed Ni atoms act as point defect scattering centers that efficiently scatter intermediate-and high-frequency phonons. Unlike crystalline defects, these Ni intercalants scatter phonons independently of lattice symmetry or periodicity due to their disordered arrangement. Second, the newly formed Ni-Se chemical bonds introduce localized structural distortions that perturb phonon propagation, significantly reducing the phonon mean free path. These combined effects lead to enhanced phonon scattering and consequently lower lattice thermal conductivity in the Ni-intercalated compound.

Figs. 5b and 5e present the temperature-dependent Seebeck coefficients of $In_2Se_3$ and $Ni_{0.3}In_2Se_3$. Comparative analysis shows that the anisotropy of the Seebeck coefficient is less than that of electrical and thermal conduction. Figs. 5c and 5f shows the temperature-dependent thermoelectric figure of merit ($ZT$) of $In_2Se_3$ and $Ni_{0.3}In_2Se_3$. After Ni intercalation, thermoelectric performance shows significant enhancement. The in-plane $ZT$ value increases from 0.2002 to 0.3716, representing a substantial improvement of 86%. The out-of-plane $ZT$ value rises from 0.2761 to 0.8896, demonstrating an exceptional enhancement of 222%. The more pronounced improvement in out-of-plane thermoelectric performance compared to in-plane performance can be primarily attributed to the enhancement of the electrical conductivity and the decrease of the thermal conductivity in the out-of-plane direction. This directional difference in electrical and thermal transport modification following Ni intercalation directly translates to the anisotropic enhancement observed in the thermoelectric figure of merit. In addition, it is worth noting that both the electrical conductivity and thermoelectric properties of $In_2Se_3$ change rapidly with increasing temperature, with a particularly dramatic change at ~200 °C due to the phase transition. This fact, to some extent, may lead to unexpected result when $In_2Se_3$ is used for high-temperature applications. In contrast, the temperature-dependent properties of $Ni_{0.3}In_2Se_3$ are more steady and predictable. Thus, the significance of Ni intercalation is demonstrated by its ability to not only enhance the thermoelectric performance of $In_2Se_3$ but also improve its viability for high-temperature applications.

## Conclusion

In this work, an unconventional ternary compound $Ni_xIn_2Se_3$ was synthesized through an intercalation route. The maximum intercalation concentration x is determined as 0.3 through XRD, SEM and EDS analysis. In $Ni_xIn_2Se_3$, the Ni atoms randomly occupy the tetrahedral sites within the $In_2Se_3$ interlayer. The disordered intercalation of Ni significantly enhances the electrical conductivity of $In_2Se_3$ while decreasing the thermal conductivity, thus improving the thermoelectric properties. Compared with $In_2Se_3$, the thermoelectric preferences (*ZT*) of $Ni_{0.3}In_2Se_3$ at 500 °C increased by 86% (in-plane) and 222% (out-of-plane). Meanwhile, the Ni intercalation effectively suppresses the phase transition of $In_2Se_3$ at high temperature. When the temperature increases from room temperature to 250 °C, $In_2Se_3$ undergoes an α-to-β-to-γ phase transformation, leading to a dramatic change in physical properties. In contrast, $Ni_{0.3}In_2Se_3$ remains stable in the β-phase across the entire temperature range (room temperature to 500 °C), and exhibits stable and linearly temperature-dependent physical properties.

**Acknowledgements**

This study was financially supported by the National Natural Science Foundation of China (Grant No. 52172254, U23A2093, and 12435017), Ningbo Natural Science

Foundation (Grant No. 20221JCGY010746), Leading Innovative and Entrepreneur Team Introduction Program of Zhejiang (Grant No. 2019R01003), Key R & D Projects of Zhejiang Province (Grant No. 2022C01236). Mian Li also acknowledges the supporting of Youth Innovation Promotion Association CAS (Grant No. 2022298).**References**

# Supporting information
# Phase stabilization of In$_2$Se$_3$ by disordered Ni intercalation and its enhanced thermoelectrical performance

Zengguang Shi[1,2#], Yukun Xiao[2,3#], Mian Li[2,3*], Jianfeng Cai[2], Yanmei Chen[2,3], Jun Jiang[2], Xiaoping Ouyang[4], Zhifang Chai[2,3], Qing Huang[2,3*]

[1] School of Materials Science and Chemical Engineering, Ningbo University, Ningbo, Zhejiang 315211, China

[2] Ningbo Institute of Materials Technology and Engineering Chinese Academy of Sciences, Ningbo 315201, China

[3] Qianwan Institute of CNITECH, Ningbo 315336, China.

[4] School of Materials Science and Engineering, Xiangtan University, Xiangtan 411105, China

E-mail addresses: limian@nimte.ac.cn (M. Li), huangqing@nimte.ac.cn (Q. Huang)

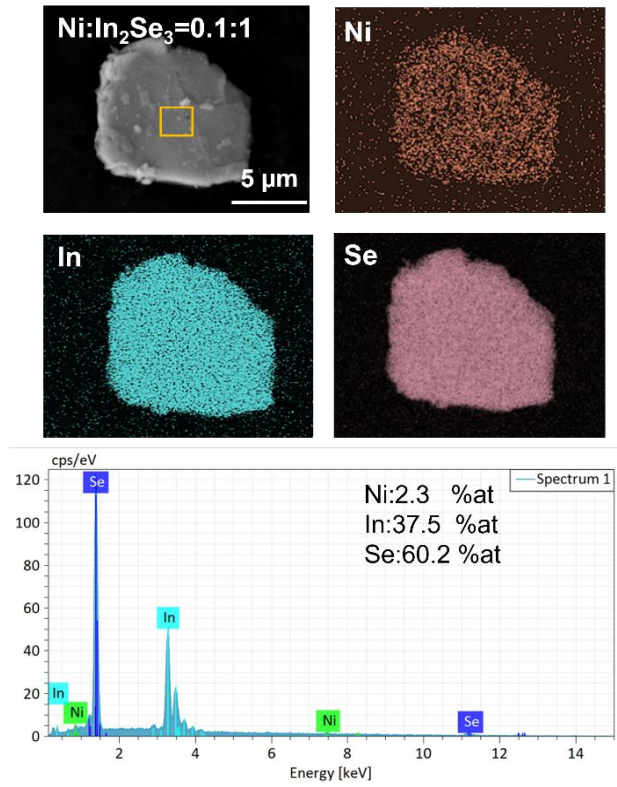

Fig. S1 SEM and EDS spectra of the intercalation products at Ni:In$_2$Se$_3$=0.1:1

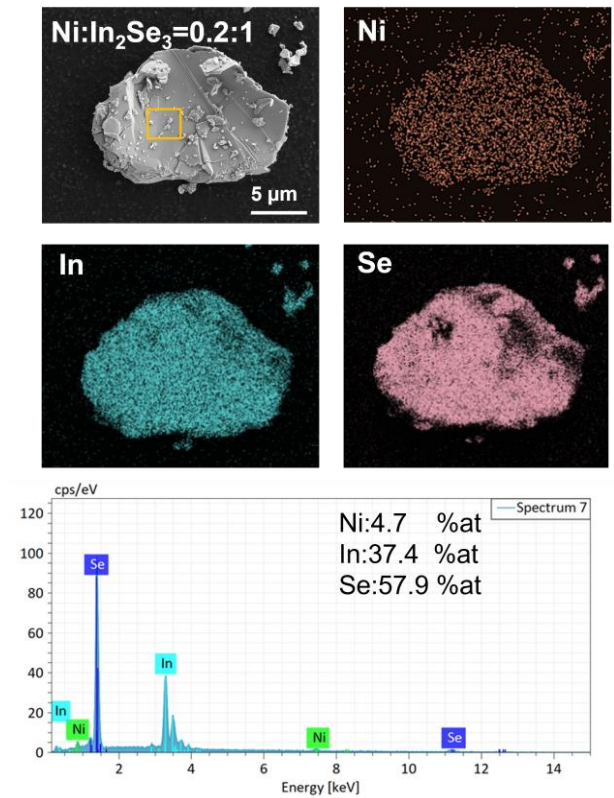

Fig. S2 SEM and EDS spectra of the intercalation products at Ni:In$_2$Se$_3$=0.2:1

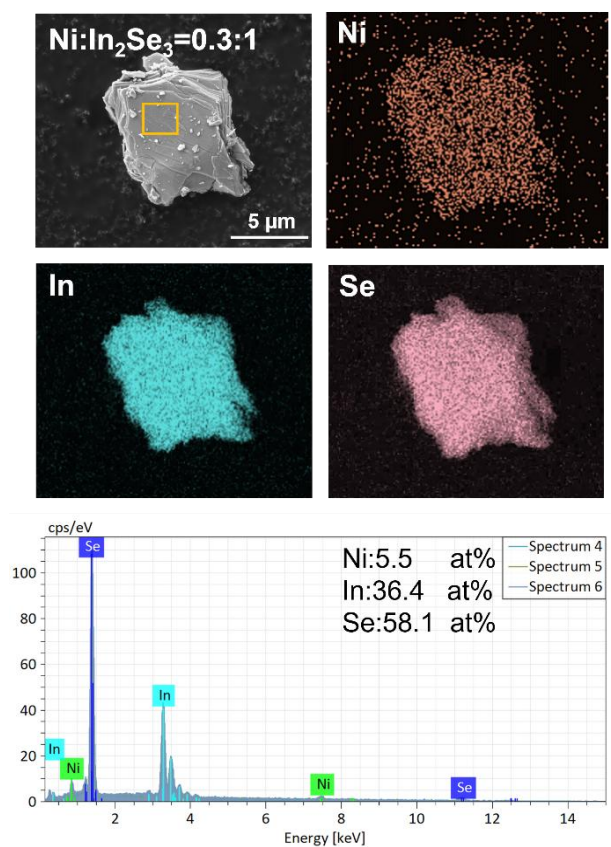

Fig. S3 SEM and EDS spectra of the intercalation products at Ni:In$_2$Se$_3$=0.3:1

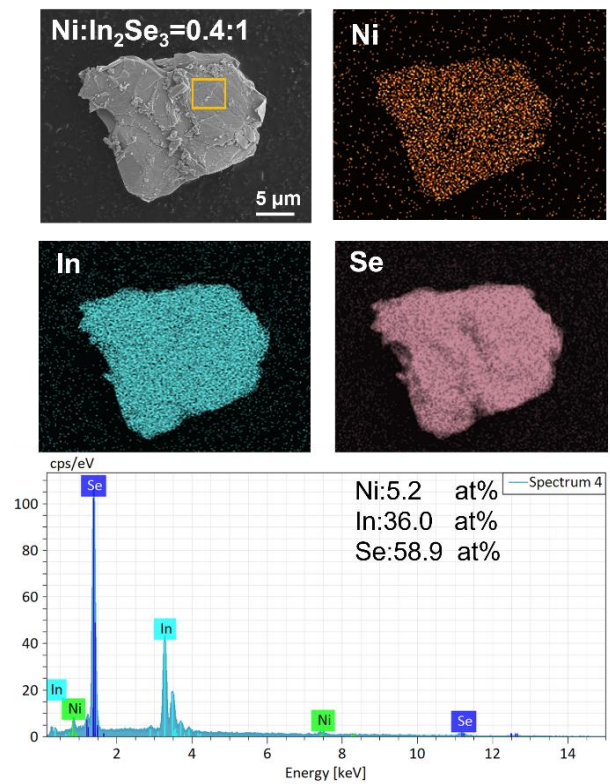

Fig. S4 SEM and EDS spectra of the intercalation products at Ni:In$_2$Se$_3$=0.4:1

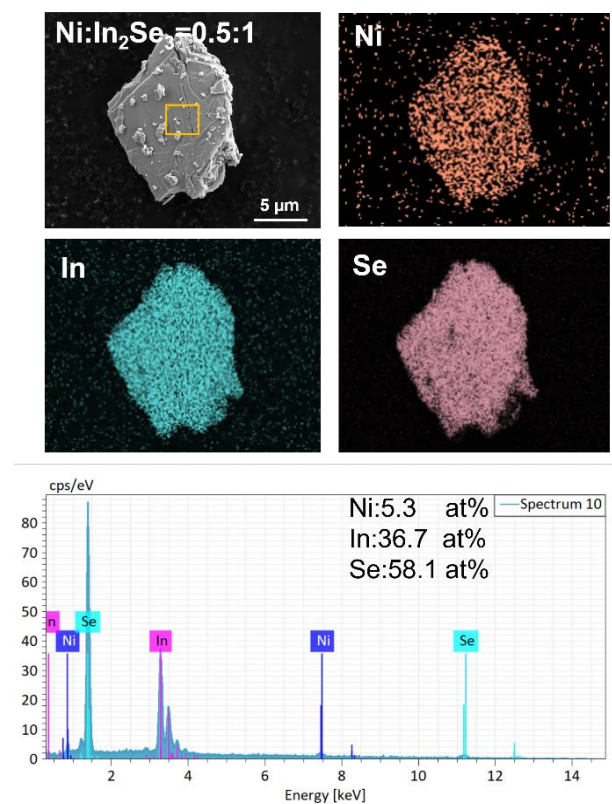

Fig. S5 SEM and EDS spectra of the intercalation products at Ni:In$_2$Se$_3$=0.5:1

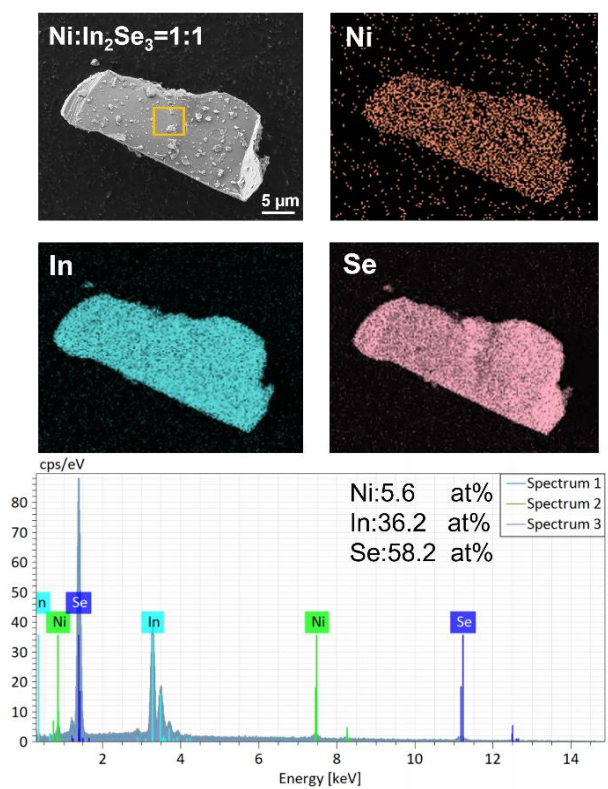

Fig. S6 SEM and EDS spectra of the intercalation products at Ni:In$_2$Se$_3$=1:1

Table S1 Summary of EDS spectral data for Ni:In$_2$Se$_3$=X:1

| Ni:In$_2$Se$_3$ | Elemental ratios (at%) | | |
| --- | --- | --- | --- |
| | Ni | In | Se |
| 0.1 | 2.3 | 37.5 | 60.2 |
| 0.2 | 4.7 | 37.4 | 57.9 |
| 0.3 | 5.5 | 36.4 | 58.1 |
| 0.4 | 5.2 | 36.0 | 58.9 |
| 0.5 | 5.3 | 36.7 | 58.1 |
| 1 | 5.6 | 36.2 | 58.2 |

Table S1 shows that at Ni:In$_2$Se$_3$ less than 0.3, the Ni content of the insertion product increases linearly with the increase of the starting Ni content, and the Ni content of the insertion product no longer increases when at Ni:In$_2$Se$_3$ greater than 0.3, at which point the excess Ni generates the Ni$_3$In$_2$Se$_2$ compound impurity. This phenomenon proves that there is a maximum value of Ni insertion content and the maximum value

is 0.3.

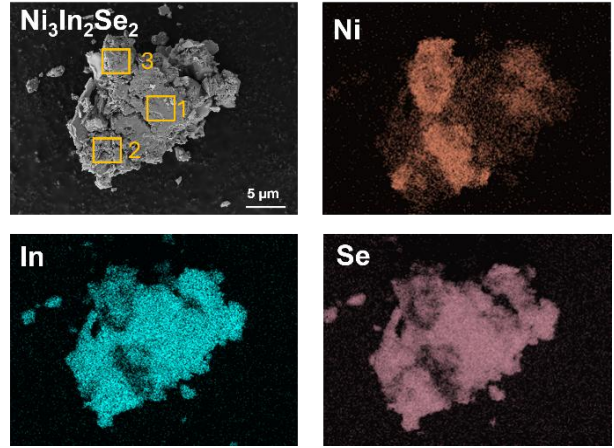

Fig. S7 SEM and EDS energy spectra of $Ni_3In_2Se_2$ compounds in $Ni:In_2Se_3$=1:1

Table S2 $Ni_3In_2Se_2$ Elemental content

| | Elemental content (at%) | | |
|---|---|---|---|
| | Ni | In | Se |
| 1 | 9.6 | 36.6 | 53.8 |
| 2 | 28.2 | 27.2 | 44.1 |
| 3 | 62.9 | 22.2 | 14.9 |

Fig. S7 and Table S2 shows the process of destroying the structure of $In_2Se_3$ by excess Ni, the Ni atoms gradually diffuse from the outside to the inside of $In_2Se_3$, and the entry of excess Ni directly destroys the crystal structure of $In_2Se_3$, which transforms the crystal structure from a layered structure to an irregular structure.

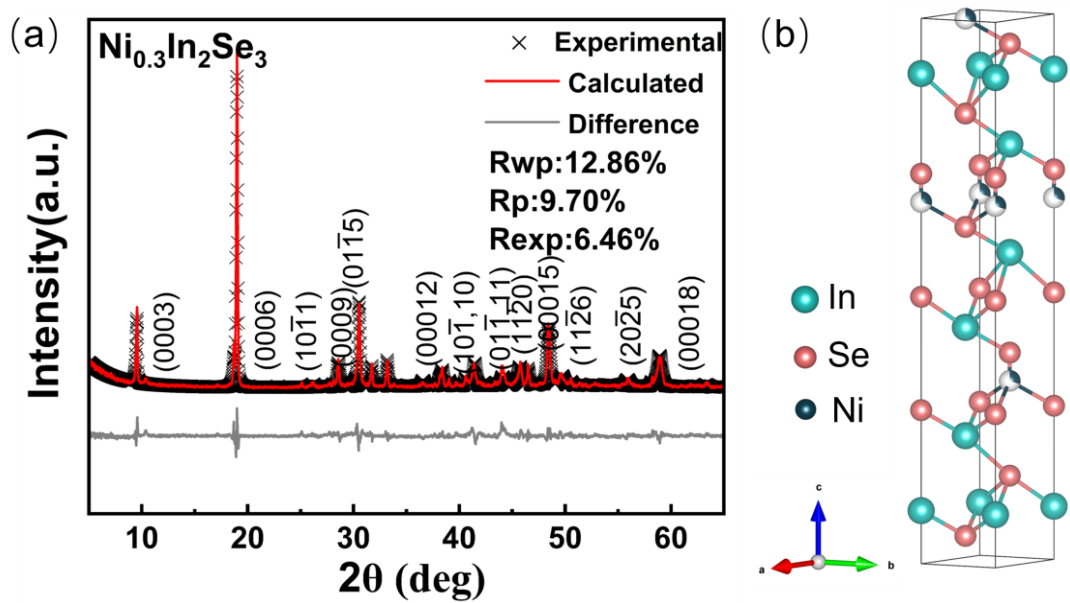

Fig.S8 Refined XRD pattern and structure of Ni$_{0.3}$In$_2$Se$_3$

Table S3 Atomic positions of Ni$_{0.3}$In$_2$Se$_3$

| Sample | a/nm | c/nm | Element | x | y | z | Occupied |
|---|---|---|---|---|---|---|---|
| Ni$_{0.3}$In$_2$Se$_3$ | 0.3973 | 2.831 | In1 | 2/3 | 1/3 | 0.2321 | 1 |
| | | | In2 | 2/3 | 1/3 | 0.43081 | 1 |
| | | | Se1 | 2/3 | 1/3 | 0.04846 | 1 |
| | | | Se2 | 1/3 | 2/3 | 0.15617 | 1 |
| | | | Se3 | 0 | 0 | 0.28164 | 1 |
| | | | Ni | 1/3 | 2/3 | 0.34050 | 0.3 |

Table S3 Atomic positions in Ni$_{0.3}$In$_2$Se$_3$ determined form the Rietveld refinement of the XRD pattern shown in Fig. S8. The space group was *R3m* symmetry. The *a* and *c*-lattices parameters of Ni$_{0.3}$In$_2$Se$_3$ were calculated to be 3.973 Å and 28.31 Å. The reliability factors after Rietveld refinement are Rwp = 12.86% and Rp = 9.70%.

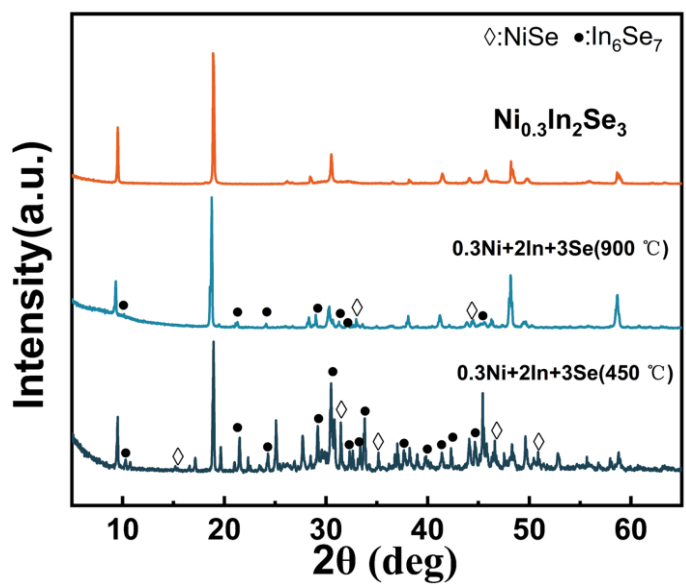

Fig.S9 Solid-phase reaction direct sintering of Ni$_{0.3}$In$_2$Se$_3$ XRD diffractograms

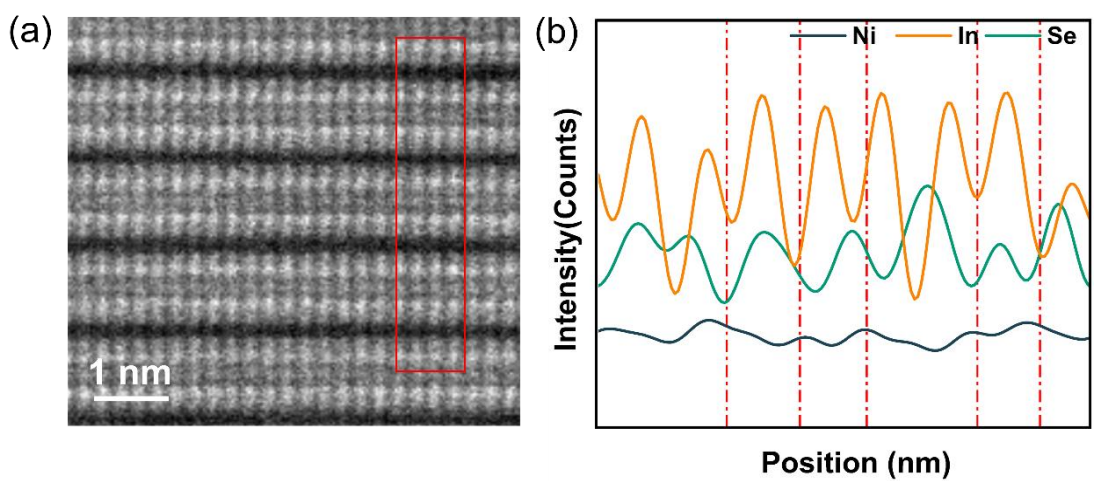

Fig.S10 HAADF-STEM images and their corresponding line scans

Fig S10a shows the HADDF-STEM image of $Ni_{0.3}In_2Se_3$, and Fig. b shows its corresponding line-scanning energy spectrum. The line-scanning energy spectrum observed that the peaks of the energy spectrum of Ni are located in the peaks and valleys of the In and Se elements, which indicates that Ni is in the interlayer region where there are no In and Se elements, which is a further proof of the insertion of Ni.

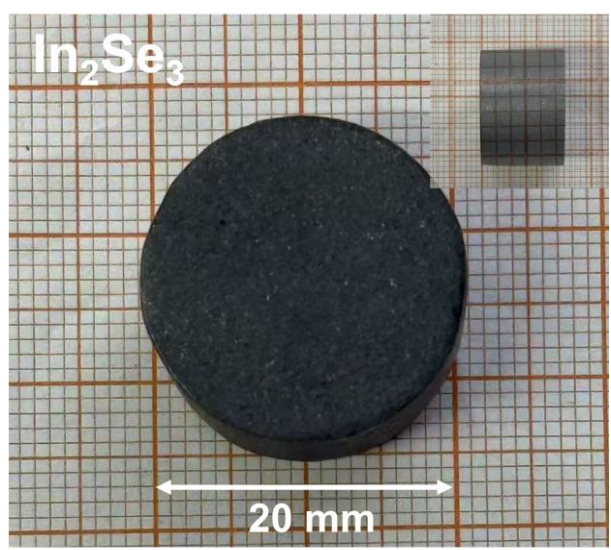

Fig S11 $In_2Se_3$ bulk sample topography

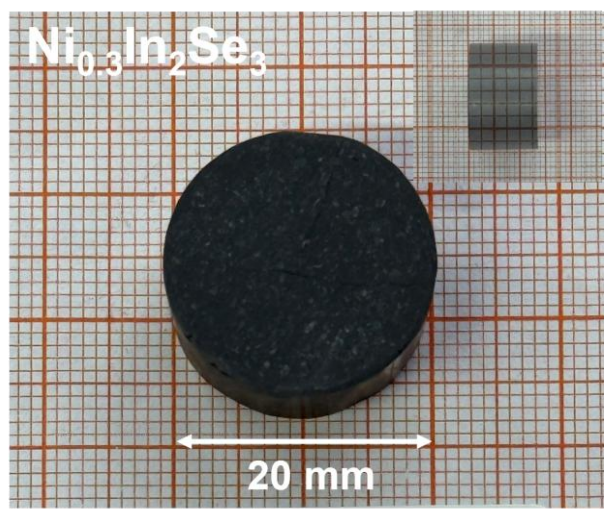

Fig S12 $Ni_{0.3}In_2Se_3$ bulk sample topography

Table S4 Sample Density Measurement

| Sample | Actual density | Rheoretical density | Relative density |
|---|---|---|---|
| $In_2Se_3$ | 5.68g/cm³ | 5.7583 g/cm³ | 98.6% |
| $Ni_{0.3}In_2Se_3$ | 6.01g/cm³ | 6.254g/cm³ | 96.1% |

The theoretical densities of $In_2Se_3$ and $Ni_{0.3}In_2Se_3$ in Table S4 were obtained from the atomic structure diagrams given in Table S3 and Table S5.

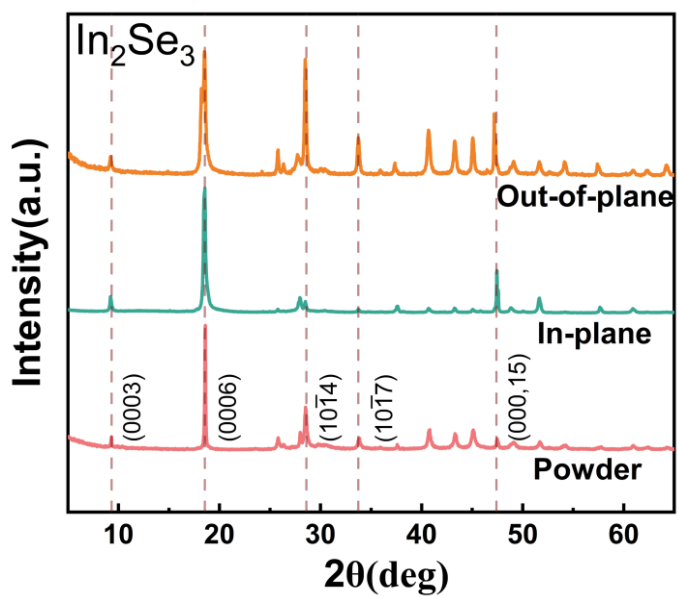

Fig S13 XRD patterns of $In_2Se_3$ powder and bulk simples cut along different directions.

Table S5 Atomic positions of α-$In_2Se_3$($R3m$)

| Sample | a/nm | c/nm | Element | x | y | z | Occupied |
|---|---|---|---|---|---|---|---|
| $In_2Se_3$ | 0.402 | 2.875 | In1 | 2/3 | 1/3 | 0.140 | 1 |
| | | | In2 | 2 | 1 | 0.328 | 1 |

|     | /3 | /3 | 900   |   |
| --- | --- | --- | --- | --- |
| Se1 | 2  | 1  | 0.053 | 1 |
|     | /3 | /3 | 800   |   |
| Se2 | 1  | 2  | 0.181 | 1 |
|     | /3 | /3 | 400   |   |
| Se3 | 0  | 0  | 0.282 | 1 |
|     |    |    | 700   |   |